\documentclass[preprint]{elsarticle}
\usepackage{amsmath}
\usepackage{amsthm}
\usepackage{amssymb}
\usepackage{amscd}
\usepackage{amsfonts}
\usepackage{amsbsy}
\usepackage{graphicx}
\usepackage{epstopdf}
\usepackage{amssymb,latexsym}
\begin{document}

\journal{Physica D}

\title{The stability of the triangular libration points for the plane circular restricted three-body problem with light pressure }

\author[UAM]{M. Alvarez-Ram\'{\i}rez}
\ead{mar@xanum.uam.mx}
\address[UAM]{Departamento de Matem\'aticas, UAM-Iztapalapa, San Rafael Atlixco 186, Col. Vicentina, 09340 Iztapalapa, M\'exico, D.F., M\'exico. }

\author[FATEC]{J. K. Formiga}
\ead{jkennety@yahoo.com.br}
\address[FATEC]{FATEC - Faculty of Technology, 12247-014,  S\~{a}o Jos\'e dos Campos  SP, Brazil.}

\author[UNIFESP]{R. V. de Moraes}
\ead{rodolpho@feg.unesp.br}
\address[UNIFESP]{UNIFESP-Univ Federal de S\~{a}o Paulo, 12231-280, S\~{a}o Jos\'e dos Campos  SP, Brazil.}

\author[UERJ]{J. E. F. Skea}
\ead{jimsk@dft.if.uerj.br}
\address[UERJ]{Departamento de F\'{\i}sica Te\'orica, Universidade do Estado do Rio de Janeiro, 20550-900, Rio de Janeiro RJ, Brazil.}

\author[UFRJ]{T. J. Stuchi}
\ead{tstuchi@if.ufrj.br}
\address[UFRJ] {Universidade Federal do Rio de Janeiro, 21941-909, Rio de Janeiro RJ,  Brazil.}



\date{}

\maketitle

\begin{abstract}We study the fourth-order stability of the triangular libration points in the absence of resonance for the
three-body problem when the infinitesimal mass is affected not only by gravitation but also by light pressure from both primaries.
A comprehensive summary of previous results is given, with some inaccuracies being corrected. The Lie triangle method is used to
obtain the fourth-order Birkhoff normal form of the Hamiltonian,  and the corresponding complex transformation to pre-normal
form is given explicitly. We obtain an explicit expression for the determinant required by the Arnold-Moser theorem, and show that it
is a rational function of the parameters, whose numerator is a fifth-order polynomial in the mass parameter. Particular cases where
this polynomial reduces to a quartic are described. Our results reduce
correctly to the purely gravitational case in the appropriate limits, and extend numerical work by previous authors.
\end{abstract}

\section{Introduction}
Stars (including the Sun) exert not only gravitation, but also radiation pressure on nearby bodies.
At the same time, it is well known that dust particles are characterized by a considerable sailing capacity
(cross-section to mass ratio), and hence are subject to a sizable effect of light pressure from the star,
being one of the possible mechanisms for the formation and evolution of gas-dust clouds.

In the classical planar restricted circular three-body problem, two large masses,
$m_1$ and $m_2$, rotate in planar circular Keplerian orbits, while
a third  particle of negligible mass moves in the same plane as the
two larger bodies under their gravitational pull. However the classical model of the
restricted three-body problem is not valid for studying the motion of material points
in the solar system where the third mass has considerable sailing capacity
(for example cosmic dust, stellar wind, etc).
Thus it is reasonable to modify the classical model by superposing a radiative repulsion field,
whose source coincides with the source of the gravitational field (the Sun), onto the gravitational field of the main body.

This problem  was called the {\em photo-gravitational restricted three-body problem} by Rad\-zi\-ev\-skii~\cite{rad}.
In a later work~\cite{rad53}, Radzievskii performed a complete treatment of the behavior of the equilibrium points.
In both papers, however, Radzievskii, who was primarily interested in the solar problem, only treated
a limited range of radiation pressure parameters (in particular when only one massive
body is luminous) and did not consider the question of the linear stability
of the equilibrium points.

The  photogravitational restricted three-body problem has been treated by several authors.
In 1970 Chernikov~\cite{cher} investigated the stability of the collinear equilibrium points $L_1$, $L_2$, and $L_3$,  as well as
the $L_4$ and $L_5$ points, and discussed the modifications of the results
brought about through the Poynting-Robertson effect, but again only for the Sun-planet problem.
Later, in 1985, Simmons et al.~\cite{sim} investigated the existence and linear stability of the libration points.
while, in 1996, Khasan~\cite{kh} studied librational solutions of the photogravitational restricted three body problem by considering both primaries as radiating bodies.

Nonlinear stability of the triangular libration points was investigated by Kumar and Choudhry~\cite{ku},
who extended the work of  Radzievskii by analyzing the stability of the triangular points for all values of the parameters
which describe the radiating effects of the primaries. They found that, except for some cases, the motion is stable for all
values of the radiation reduction factors and for all values of $\mu < 0.0285954\ $.
They also studied~\cite{ku88}
the stability of $L_4$ and $L_5$ under the resonance conditions $\omega_1=2\omega_2$ and $\omega_1=3\omega_2$.
Later, Go\'zdziewski et al.~\cite{gmn} also studied the nonlinear stability of the triangular libration points. Their study of the
stability of the libration points when the resonances do not exist, and in the fourth
order resonance case, shows that for some values of parameters these points are
stable, and for others they are unstable. In the case when the third order resonances
exist, the triangular libration points are always unstable.

Kunitsyn and Polyakhova~\cite{ku-po} gave a retrospective review of many aspects of the
libration point positions and their stability for all values of radiation pressure and mass ratios.

In this paper we investigate the Lyapunov stability of the triangular solutions for all possible values of the reduction coefficients given by $\kappa_1$ and $\kappa_2$,
which represent the ratios of the difference between the gravitational force and the radiative force of the bodies of mass $m_1$ and $m_2$, respectively.

We would like to stress here that some results presented in this paper have been obtained in the above references.
The results of this paper should  be viewed as a step towards describing the  dynamics of the photogravitational restricted three-body problem.
We note that, seen as a whole, the problem displays a number of interesting features that are not apparent in previous treatments.

The paper is organized as follows: section~\ref{sec2}  contains preliminaries, where
we recall the earlier results on the photogravitational planar restricted three body problem, and introduce  the  Hamiltonian function and the equilibria of the
system, along with the condition on the masses and radiation pressure values
for the stability of the linearized problem.

In Section~\ref{sec3}, the study of the existence and linear stability of equilateral libration points is presented.

Finally, in  Section~\ref{sec-normal},
we apply the Arnold-Moser theorem to examine the condition of non-linear stability, excluding the resonance cases up to the fourth order.
We obtain a condition for stability in the parameter space of $\kappa_1$, $\kappa_2$ and $\mu$ which may be expressed as a quintic in $\mu$,
plot the relevant surface, and show that this polynomial reproduces the classical result of
Deprit and Deprit-Bartholom\'{e}~\cite{Deprit} when $\kappa_1=\kappa_2=1$. Special cases when this polynomial reduces to a quartic (and therefore
explicit algebraic expressions for the surface may be obtained) are determined.


\section{Hamiltonian formulation}\label{sec2}
We consider an infinitesimal mass particle moving in the  photogravitational field of two masses, termed the {\em primaries}, $m_1$ and $m_2$,
with both masses in circular orbits around their common center of mass.
The two primaries are sources of radiation, with the parameters $\kappa_1$ and $\kappa_2$ characterizing the radiation effect of $m_1$ and $m_2$ respectively.
Similar to the planar classical case, the motion of the infinitesimal mass takes place in the same plane of the primaries.

We should note that, in contrast to the classical restricted three-body problem, in the photogravitational  problem the force acting on the particle depends not only on
the parameters of the stars (temperature, size, density, etc.) but also on the parameters of the particle itself (size, density, etc.)

The photogravitational version of the restricted problem presented here is derived in a similar way to the classical problem (see~\cite{szebe}): in fact we start from a presentation and
notation very similar to that used by Simmons~\cite{sim}.
Units are chosen so that the unit of mass is equal to the sum of the primary masses, $m_1 + m_2=1$, the unit of length is equal to their separation,
and the unit of time is such that the angular velocity $\omega=1$. We also set the gravitational constant $G = 1$.
For definiteness we also take  $\mu = m_2/(m_1+m_2)$, $0\leq \mu \leq 1$ so that $m_1=1-\mu$ and $m_2 =\mu$.

We have fixed the center of mass at $(0, 0)$ and the primaries, $m_1$  and $m_2$, at
$(-\mu, 0)$ and $(1 -\mu, 0)$, respectively.
The forces experienced by a test particle
in the  coordinate system rotating with $\omega=1$ and origin at the center of mass  are then derivable from the potential
\begin{equation}\label{potential}
U(x,y)= \frac{1}{2}(x^2 + y^2)+ \frac{\alpha (1-\mu)}{r_1}+ \frac{\beta \mu}{r_2},
\end{equation}
where $(x,y)$ are the coordinates of the test particle,
$$r_1^2= (x+\mu)^2 + y^2, \qquad r_2^2= (1-x-\mu)^2 + y^2$$
are the distances from the masses $m_1$ and $m_2$, respectively,
and $\alpha$, $\beta$ represent the effects of the radiation pressure from the two primaries.

The Jacobi constant of the problem is given by
\begin{equation}\label{ham1}
C_J(x,y,\dot{x},\dot{y})=\frac{1}{2}(\dot{x}^2+\dot{y}^2) + \frac{1}{2}(x^2 + y^2)+ \frac{\alpha (1-\mu)}{r_1}+ \frac{\beta \mu}{r_2}.
\end{equation}
After introducing the canonical coordinate system $(x, y, p_x, p_y)$
\begin{eqnarray*}
p_x=\frac{dx}{dt}-y, \quad  p_y=\frac{dy}{dt}+x
\end{eqnarray*}
one obtains the Hamiltonian function
\begin{equation}\label{ham2}
H(x,y,p_x,p_y)= \frac{1}{2}(p_x^2+p_y^2)+p_xy-p_yx+ \frac{\alpha (1-\mu)}{r_1}+ \frac{\beta \mu}{r_2}.
\end{equation}

In this canonical formulation the problem of determining the libration points consists of finding  all real solutions of the system of four algebraic equations
given by setting the Hamiltonian equations of motion equal to zero, that is
\begin{align}\label{lib-point}
& p_x+y=0, \qquad p_y-x=0,  \nonumber \\
-&p_y + \frac{\alpha(1-\mu)(x+\mu)}{r_1^3}+ \frac{\beta\mu(x-1+\mu)}{r_2^3}=0, \\
& p_x +  \left( \frac{\alpha (1-\mu)}{r_1^3}+ \frac{\beta\mu }{r_2^3}\right)y=0. \nonumber
\end{align}

For later convenience we introduce the parameters $\kappa _1^3=\alpha$  and $\kappa_2^3=\beta$
as used by Schuerman~\cite{sch}.
Defining
\begin{equation}
b\equiv\displaystyle{1-\left(\frac{\kappa_1^2+\kappa_2^2-1}{2\kappa_1 \kappa_2}\right)^2},
\label{bdef}
\end{equation}
we find that the solution of~(\ref{lib-point}) for the
coordinates of the triangular libration points $L_4$ and  $L_5$ is~(see~\cite{ku}):
\begin{align}\label{lib-coor}
& x_{L_4} = x_{L_5} = \displaystyle{\frac{\kappa_1^2+1-\kappa_2^2}{2}-\mu}, \qquad
p_{x_{L_4}}= - p_{x_{L_5}} = - \kappa_1\kappa_2\sqrt{b},\nonumber\\ \\
& y_{L_4}= -y_{L_5}= \kappa_1\kappa_2\sqrt{b}, \quad \qquad   p_{y_{L_4}}= p_{y_{L_5}}= x_{L_4}. \nonumber
\end{align}
This implies that
\begin{equation}\label{rel}
r_1=\alpha^{1/3}= \kappa_1, \quad\quad r_2=\beta^{1/3}= \kappa_2,
\end{equation}
showing that a necessary condition for the existence of  triangular points is $\kappa_1 \geq 0$ and $\kappa_2 \geq 0$~(see~\cite{sim}).
In fact, as shown by Schuerman~\cite{sch}, the points lie at the intersections of the circles
defined by~(\ref{rel}), and so they exist provided further that $\kappa_1 + \kappa_2 \geq  1$.
We  remark that the existence  of the equilibrium solutions is critically governed by the numerical value of $b$.

These points are the vertices of two triangles, of sides $\kappa_1$, $\kappa_2$ and $1$, based on the line joining the
primaries, and so $L_4$ and $L_5$ are known as triangular (Lagrange) libration points. If we set $\alpha = 1$ and $\beta_2 = 1$, the restricted
photogravitational three body problem is reduced to the classical case. We consider here only the stability of $L_4$,
however all conclusions about the stability of $L_4$ can be extended to $L_5$ just applying the symmetries of the photogravitational problem, namely
\begin{equation}\label{simetria}
(x,y,p_x,p_y,t) \mapsto (x,-y,-p_x,p_y,t).
\end{equation}

\section{Linear stability}\label{sec3}
To investigate the stability we use Birkhoff's procedure~\cite{birkhoff} for normalizing the Hamiltonian
in a neighborhood of the libration point.
We start the study of stability by  finding the first order variational equations, which we then use to determine the eigenvalues.
By the Arnold-Moser theorem~(see~\cite{meyer}), it is known that a necessary condition for stability of the $L_4$ point is that all
eigenvalues should be pure imaginary.




Using a linear canonical transformation, we first shift the origin of the coordinate system to the $L_4$ point
and expand the Hamiltonian in a power series of the coordinates.
To this end we define new coordinates  $q_1$, $q_2$, $p_1$, $p_2$  by
\begin{eqnarray*}
x=x_{L_4} + q_1, \quad && \quad p_x= p_{x_{L_4}} + p_1, \\
y=y_{L_4} + q_2, \quad && \quad p_y= p_{y_{L_4}} + p_2,
\end{eqnarray*}
in terms of which the equations of motion are
$$
\frac{dq_i}{dt}= \frac{\partial H}{\partial p_i}, \qquad
\frac{dp_i}{dt}= -\frac{\partial H}{\partial q_i}, \qquad  i=1,2.
$$
In these new variables the solution (\ref{lib-coor}) corresponds to the equilibrium state
$q_j=p_j=0$, for  $j=1,2$.F
We now expand the Hamiltonian in a power series of the coordinates, up to fourth order.
Since the expansion is made in the neighborhood of an equilibrium point, the constant term (the value
of the Hamiltonian at equilibrium) can be neglected, and the linear part must vanish.
The expanded Hamiltonian can be written as
\begin{equation}\label{ham-formal}
H= \sum_{j=0}^\infty H_j, \end{equation}
where $H_j$ are homogeneous polynomials of degree $j$ in the new variables.
Calculating this expansion to fourth order we find
\begin{align*}
H_0 &= H(x_{L_4},y_{L_4},p_{x_{L_4}},p_{y_{L_4}}), \\
H_1 & =0,\\
H_2 &= \tfrac{1}{2}(p_1^2+ p_2^2) + q_2p_1-q_1p_2 + \left(A-\tfrac{1}{4}\right) q_1^2 + Bq_1q_2 -\left(A+\tfrac{1}{4}\right) q_2^2,\\
H_3 &=  h_{3000}\; q_1^3+  h_{2100}\; q_1^2q_2 + h_{1200}\; q_2^2 q_1 +h_{0300}\; q_2^3,\\
H_4 &= h_{4000}\;q_1^4+  h_{3100}\; q_1^3q_2 + h_{2200}\; q_2^2 q_1^2 +h_{1300}\; q_1q_2^3  +h_{0400}\; q_2^4,
\end{align*}
where
\begin{eqnarray}
A  & \equiv & -\frac{3}{4} +\frac{3b}{2} \left[\mu\kappa_1^2 +(1-\mu)\kappa_2^2\right], \nonumber\\[3mm]
B & \equiv & -\frac{3}{2}\,\frac{\sqrt{b}}{\kappa_1\kappa_2}\left[(1-\mu)\kappa_2^2 (\kappa_1^2+1 -\kappa_2^2)+\mu\kappa_1^2(\kappa_2^2+1 -\kappa_1^2)\right].
\label{ABdef}
\end{eqnarray}
The values of $A$ and $B$ used in  $H_2$ are conveniently defined for simplifying expressions which appear later.
The coefficients of $H_3$ and $H_4$, calculated using Maple, coincide with
those given in Kumar and Choudhry~\cite{ku} and are listed in Appendix~A.


The system of linear differential equations derived from the quadratic term, $H_2$,
describes the tangent flow around $L_4$. Linear stability is determined by
the character of the associated eigenvalues, which are the roots of the characteristic
equation
\begin{equation}\label{eq-char}
\lambda^4+\lambda^2 + 9\mu(1-\mu)\,b=0,
\end{equation}
whose eigenvalues are found to be
\begin{align}\label{eigen}
\lambda_{1,2} &= \pm \sqrt{-\frac{1}{2}+ \frac{\sqrt{1-36\mu(1-\mu)b}}{2}}\; , \nonumber\\
& \\
\lambda_{3,4} &= \pm \sqrt{-\frac{1}{2}- \frac{\sqrt{1-36\mu(1-\mu)b}}{2} }\; . \nonumber
\end{align}
For linear stability all of these eigenvalues should be pure imaginary, which is the case if
\begin{equation}\label{eq-cond}
0\leq 36\mu(1-\mu)b \leq 1.
\end{equation}
Substituting~(\ref{bdef}) in~(\ref{eq-cond}), we obtain
\begin{equation}
1-\left(\frac{\kappa_1^2+\kappa_2^2-1}{2\kappa_1 \kappa_2}\right)^2 \leq \frac{1}{36\mu(1-\mu)}
\Rightarrow \left(\frac{\kappa_1^2+\kappa_2^2-1}{2\kappa_1 \kappa_2}\right)^2 \geq \frac{-36\mu^2+36\mu -1}{36\mu(1-\mu)}.
\label{eq-cond-geral}
\end{equation}
Since the left-hand side can be zero, we require
\begin{equation}\label{eq-cond1}
 \frac{-36\mu^2+36\mu -1}{36\mu(1-\mu)} \leq 0
\Rightarrow
36\mu^2-36\mu +1 \geq 0,
\end{equation}
as $\mu\in[0,1]$. This equation is satisfied for $ \mu \leq \frac{1}{2}-\frac{\sqrt{2}}{3}$ and $ \mu \geq \frac{1}{2}+\frac{\sqrt{2}}{3}$,
and so, in these subintervals, the eigenvalues~(\ref{eigen}) are distinct and pure imaginary.

Throughout this paper, we restrict ourselves to the case
$$ \mu \leq \frac{1}{2}-\frac{\sqrt{2}}{3} \approx  0.0285954 \equiv \mu^* $$
for which the linear stability conditions are fulfilled for all values of  $\kappa_1$ and $\kappa_2$.

Note that Lagrange's classical criterion for stability is recovered in the limit of no radiation pressure
($\kappa_1 \rightarrow 1$, $\kappa_2 \rightarrow 1$, $b \rightarrow 3/4$).
In this case the inequality~(\ref{eq-cond-geral}) for the existence of two pairs of pure imaginary values becomes
$$27\mu(1-\mu) \leq 1, \quad  \mbox{or equivalently} \quad \mu (1-\mu )  \leq 1/27.$$

We remark that this result implies the stability of the $L_4$ point
in the classical planar circular restricted three-body problem when the mass ratio parameter
satisfies $0< \mu < \mu_R= \frac{1}{2} (1-\frac{\sqrt{69}}{9}) \approx 0.0385208$, where $\mu_R$ is known as the Routh value.

We write the four eigenvalues as $\pm i\omega_1$, $\pm i\omega_2$ where the strictly positive numbers $\omega_1$
and $\omega_2$  ({\em frequencies}) are determined by the set of relations
\begin{equation}\label{frec}
\omega_1^2 = - \lambda_{1,2}^2= \frac{1+M}{2}, \qquad
\omega_2^2 = - \lambda_{3,4}^2= \frac{1-M}{2},
\end{equation}
where $M\equiv\sqrt{1-36\mu(1-\mu)b}$.
The expressions~(\ref{frec}) show that
\begin{equation}\label{frec2}
0 < \omega_2 < \frac{1}{\sqrt{2}} <  \omega_1 < 1.
\end{equation}
We remark that $\omega_1$ and $\omega_2$ and the coefficients of all $H_j$ are functions of the parameters $\mu$, $\kappa_1$ and $\kappa_2$.

\section{Normal form and non-linear stability}\label{sec-normal}
Having established linear stability, the next step is to transform the Hamiltonian into its Birkhoff normal form.
This normalization allows us to apply Arnold's theorem~\cite{arnold} to investigate the stability of the $L_4$ point for mass ratios $\mu < \mu^*$,
except for the resonant cases. From the coefficients of the normal form, Arnold's thorem constructs a determinant, $D$, defined later,
which, when non-zero, establishes the stability of the equilibrium point.
To obtain the Birkhoff normal form we use the Lie series method, with the calculations being performed using Maple.
To determine the domain of applicability of the results, it is necessary to obtain the resonances, which we now calculate.


\subsection{Existence of resonances}\label{reso}
In this section we study the values for which the frequencies at the equilibrium are in resonance.
Since we are working to fourth order in the normal form, we thus need to consider resonances up to fourth order
of the triangular libration points.

The stability analysis of  the $L_4$ point can be carried out if
the frequencies $\omega_1$, $\omega_2$ satisfy the non-resonance
condition
\begin{equation}\label{eq-res1}
c_1 \omega_1 + c_2 \omega_2 \neq 0
\end{equation}
for all integers $c_1$, $c_2$ such that $ |c_1| + |c_2|\leq 4$. This condition is
violated for $\omega_1=2\omega_2$ and $\omega_1=3\omega_2$, and, obviously when $\omega_1=\omega_2$.

We start the discussion by noting that the first-order resonance appears when one of the frequencies is zero.
In our case, this is possible if either $\mu=0$ or $\kappa_1+\kappa_2=1$.
For $\mu = 0$ the restricted photogravitational three body problem is reduced to
Kepler's problem in a rotating coordinate frame.  The $L_4$ point in this case is evidently
unstable in the sense of Lyapunov.

The second case has no classical equivalent: $L_4$ and $L_5$ coincide with the
inner collinear point $L_1$ (see Simmons et. al.~\cite{sim}). In particular if
$\kappa_2=1$ and $\kappa_1\rightarrow 0$ the libration points $L_4$ and $L_5$
move from their classical equilateral positions onto the
luminous mass and coalesce there with the inner libration point $L_1$.
It is clear that the motion will be unstable.

We have seen above that if  $\kappa_1 = \kappa_2 = 1$, then $\mu=\mu_R$ and the triangular
libration points are stable in the sense of Lyapunov. This agrees with the result of
Meyer Placi\'an and Yaguas~\cite{meyer3}, who demonstrated the stability of the Lagrange equilateral triangle points, $L_4$ and $L_5$, in the plane circular restricted three-body problem when the mass ratio parameter is equal to $\mu=\mu_R$, the critical value of Routh. 

\begin{figure}[ht]
\label{fig1}
\begin{center}
\includegraphics[height=44mm]{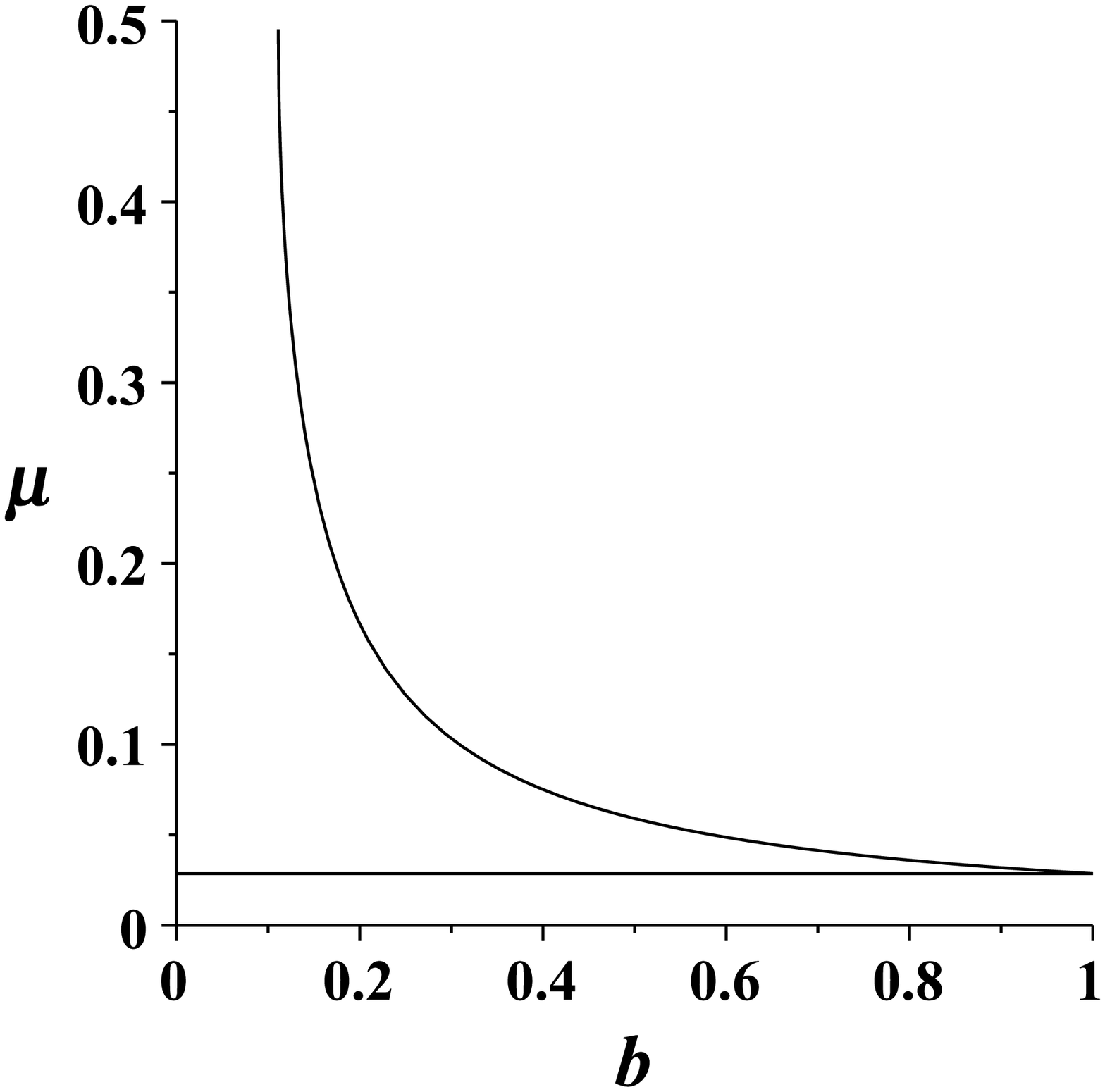}
\includegraphics[height=44mm]{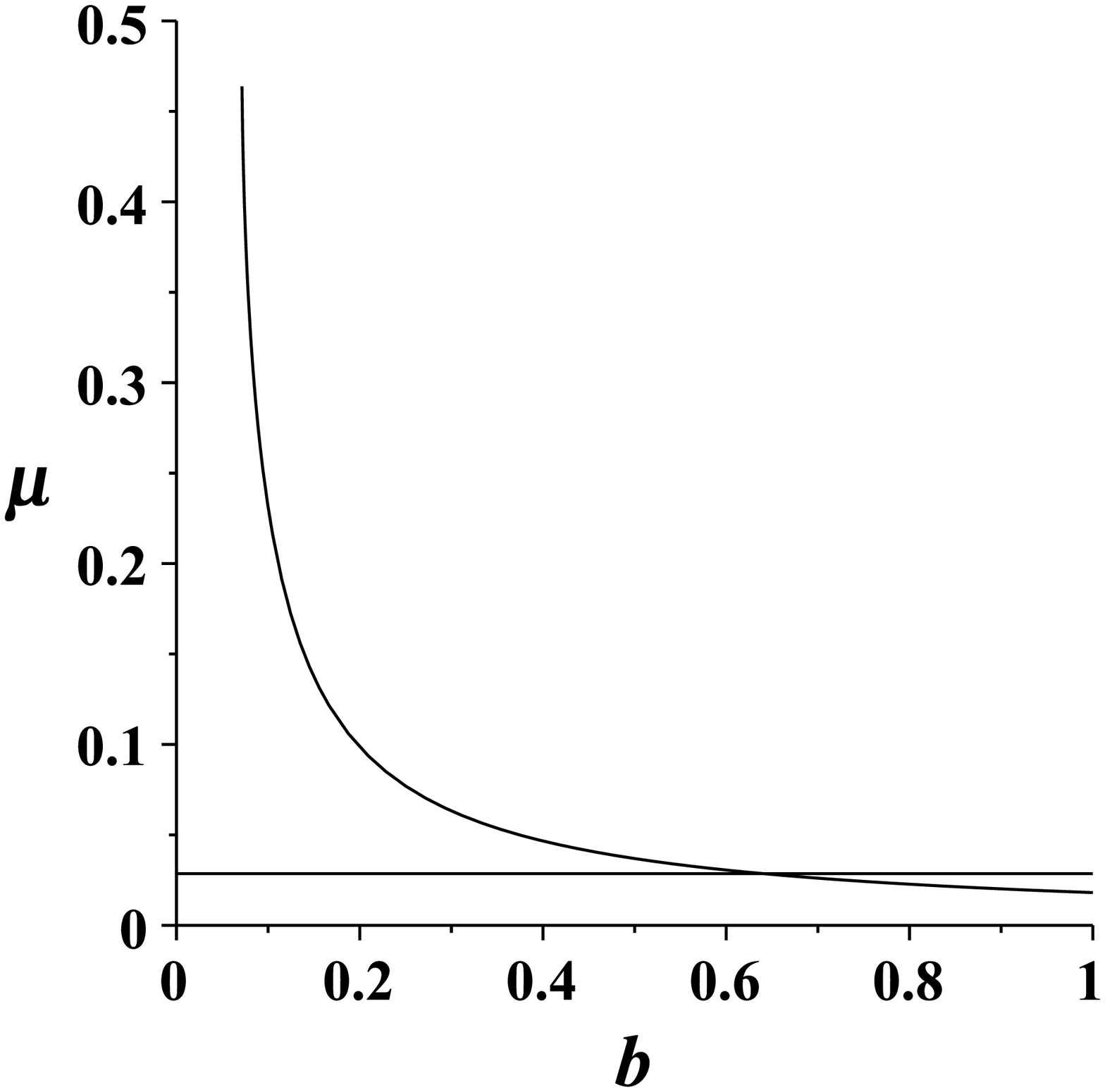}
\includegraphics[height=44mm]{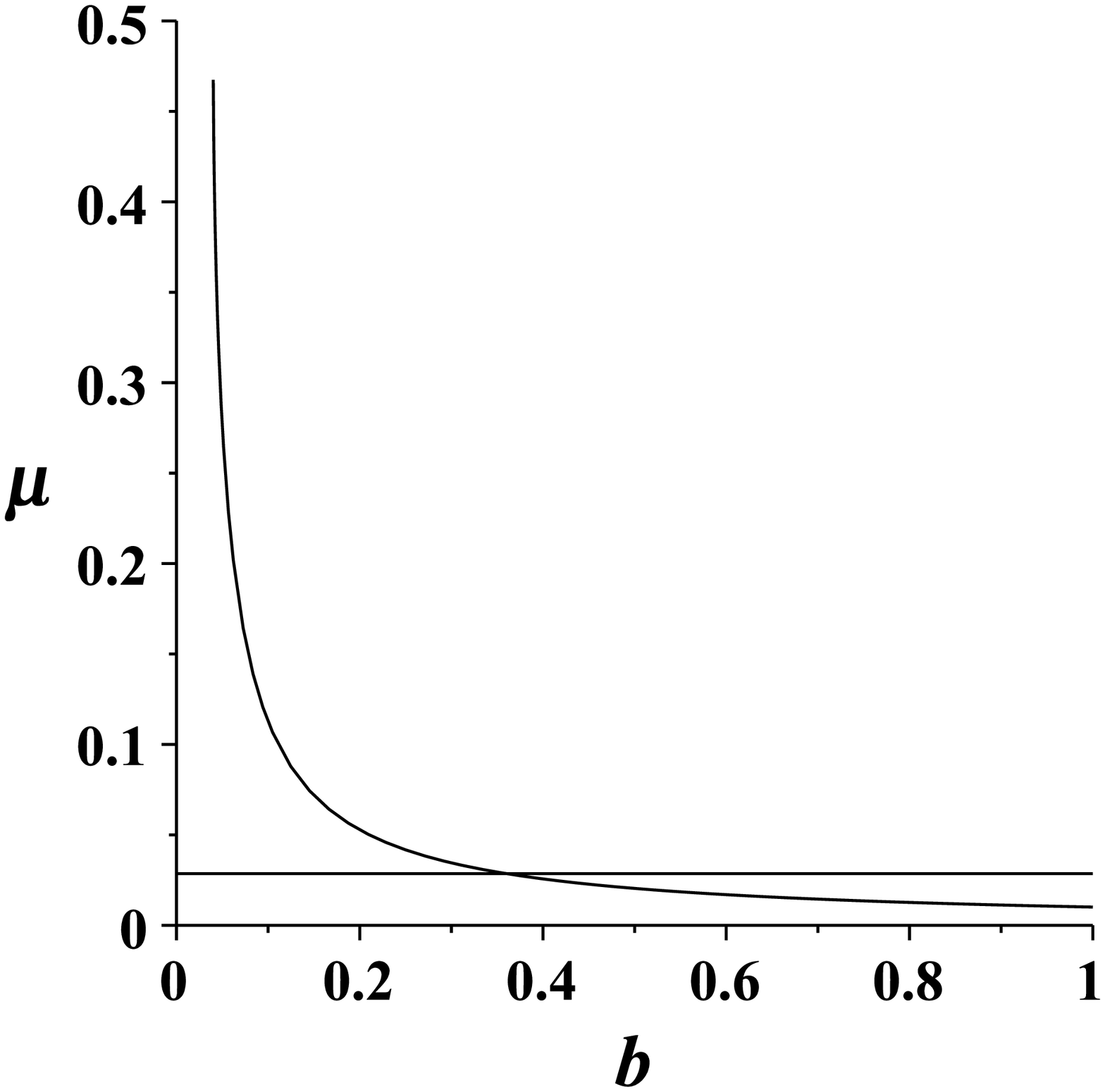}
\end{center}
\caption{Graphs of $\mu(b)$  for (from left to right) the 1:1, 1:2 and 1:3
resonances. The horizontal line is $\mu=\mu^*$.}
\end{figure}

We turn to the two remaining cases.
The second order resonance $\omega_1=\omega_2$  occurs when $\mu=\tfrac{1}{2}\pm\sqrt{\tfrac{1}{4}-\tfrac{1}{36b}}$.
Since $\mu> 0$, the resonance appears for parameters $\kappa_1$ and $\kappa_2$ satisfying the condition $b\geq 1/9$.
In the first graph of Figure~\ref{fig1} we have plotted the relevant part (i.e. with the negative root)
of $\mu(b)$, together with the line $\mu=\mu^*$.
We see that (other than at the exceptional value $b=1$) the resonant values only exist for $\mu>\mu^*$.

We use the same criterion adopted in the previous case for determining the masses for which the resonance $\omega_1=2\omega_2$ occurs.
In this way, one obtains $\mu={\frac{1}{2}-\sqrt{\frac{1}{4}-\frac{4}{225b}}}$, so that
the resonance appears for parameters $\kappa_1$ and $\kappa_2$ satisfying the condition $b\geq 16/225\approx 0.071$.
The function $\mu(b)$ for this resonance is shown in the second graph of Figure~\ref{fig1}.
We see that $\mu(b) =\mu^*$ when $b=16/25= 0.64$.
Therefore, in contrast to the results of Kumar and Choudhry~\cite{ku}, who claimed that the resonance
$\omega_1=2\omega_2$ occurs for values from $b =0.65$ to $b =0.95$, we find that it occurs for $b \in [0.64,1]$.

Finally, the resonance $\omega_1=3\omega_2$ occurs for $\mu={\frac{1}{2}-\sqrt{\frac{1}{4}-\frac{1}{100b}}}$.
Since $\mu> 0$, the resonance appears for  values of $\kappa_1$ and $\kappa_2$ satisfying ${b\geq 1/50}$.
In this case $\mu(b) =\mu^*$ for $b=9/25 = 0.36$.
Kumar and Choudhry~\cite{ku} stated that this resonance occurs for values from $b =0.4$ to $b =0.95$,
but we find that, in fact, it appears for $b \in [0.36,1]$.

In  Table~\ref{table1} we show the values of $\mu$ for $\kappa_1=1$ and several values of $\kappa_2$ for the resonances $\omega_1=\omega_2$, $\omega_1=2\omega_2$ and $\omega_1=3\omega_2$.
Note that when one of the radiation coefficients is $1$, and there is  a 1:1 resonance,  the points $L_4$ and $L_5$ are linearly stable with the condition that mass parameter $\mu$  is at most $0.0385209$.
Our result refines that of Kumar and Choudhry~\cite{ku} and has an influence on the analysis of the non-linear stability of $L_4$, and consequently on that of $L_5$.

\begin{table}
\begin{center}
\begin{tabular}{|r |l l l|}
\hline
$\kappa_2$ &$\omega_1=\omega_2$ & $\omega_1=2\omega_2$ & $\omega_1=3\omega_2$ \\
\hline
1 & 0.0385209 & 0.0349233	& 0.0321444\\
0.9 & 0.0361369	& 0.0327702	& 0.0301682\\
0.8 & 0.0342412	& 0.0310571	& 0.0285955\\
0.7& 0.0327266	& 0.029688	& 0.0273381\\
0.6 & 0.0315184	& 0.0285955	& 0.0263345\\
0.5& 0.0305638	& 0.027732	& 0.0255412\\
0.4& 0.0298247	& 0.0270634	& 0.0249269\\
0.3 & 0.0292741	& 0.0265653	 & 0.0244692\\
0.2& 0.0288932	& 0.0262206	& 0.0241524\\
0.1 & 0.0286693	& 0.0260181	& 0.0239662\\
\hline
\end{tabular}
\end{center}
\caption{\label{table1}Values of $\mu$ for $\kappa_1=1$.}
\end{table}

\subsection{Birkhoff's Normal Form}
In section~\ref{sec3} we determined the conditions under which the eigenvalues are purely imaginary,
thus guaranteeing that the phogravitational
Hamiltonian~(\ref{ham2}) satisfies the first condition of the Arnold-Moser theorem.
The next step is the normalization procedure that transforms Hamiltonian~(\ref{ham-formal}) to Birkhoff's normal form up to fourth order,
excluding regions close to the resonances established in the previous section.

To this end, first we make a further change of variables,
$$(q_1,q_2,p_1,p_2)\rightarrow ( \bar \xi_1,\bar \xi_2,\bar \eta _1,\bar \eta_2),$$ to the Hamiltonian~(\ref{ham2}) that diagonalizes $H_2$:
\begin{equation}\label{rnf2}
 \bar H =  \frac{\omega_1}{2}(\bar \xi_1^2 + \bar \eta _1^2) -\frac{\omega_2}{2}(\bar \xi_2^2 + \bar \eta _2^2) + \sum_{j=3}^m H_j( \bar \xi_1,\bar \xi_2,\bar \eta _1,\bar \eta_2) + O_{m+1}.
\end{equation}
\noindent We use a negative sign on  the second term to emphasize the non-positive character of the photogravitational Hamiltonian.

We note that $\bar H_2$ is already in real normal form since it depends only on the actions $\bar R_\nu \equiv \bar \xi_\nu^2 + \bar \eta _\nu^2$, $\nu=1,2$.
Also, recall that $\omega_1$ and $\omega_2$ are the linear frequencies analyzed
in the previous sections. The terms of order greater than two depend on $(\bar \xi_1,\bar \xi_2,\bar \eta _1,\bar \eta_2)$, but not necessarily in such
a simple way as $\bar H_2$. The purpose of the Birkhoff normal form is to find a near identity symplectic coordinate change
\begin{equation}\label{norcord}
{\bf z^{(m)}}=\bar {\bf z}+ \sum_{i+j+k+l=2}^{i+j+k+l=m} a_{ijkl} {\bar \xi_1}^i{\bar \xi_2}^j{\bar \eta_1}^k{\bar \eta_2}^l+ O_{m+1}
\end{equation}
\noindent so that the even order terms in the Hamiltonian depend on the new variables ${\bf z}=( \xi_1, \xi_2, \eta _1, \eta_2)$ in such a way
that they can be grouped as powers of $R_1$ and $R_2$.

It is implicit in this last statement that the Hamiltonian system's  natural resonances have been taken into account.
Let $\mathcal{S}$ be some system of resonance relations, {\em i.e.} relations of the form
$\lambda_s = \sum_{i=1}^n m_i\lambda_i$
with $m_i$ non-negative integers not all zero.
From the viewpoint of the general theory of dynamical systems, equilibria and periodic
trajectories in Hamiltonian systems are all resonant. Indeed, for equilibria, if $\lambda_1$ is an eigenvalue of
an equilibrium of a Hamiltonian system, then $\lambda_2 = -\lambda_1$ is also an
eigenvalue. So an infinite number of resonance relations of the form
\[
\lambda_s=\lambda_s + k(\lambda_1 + \lambda_2),  k\in\mathbb{Z}
\]
are satisfied.
This means that we can find a non-divergent generating function (and consequent canonical transformation) that
transforms away all terms of odd degree of the Hamiltonian~(\ref{ham-formal}).
We have, necessarily, to retain part of the even terms corresponding to the resonant monomials that appear in~(\ref{ham-formal}).
This will become clearer in the brief presentation of the
algorithm we have used to implement the Birkhoff normal form.
Of course, a truncated Birkhoff normal form is a polynomial Hamiltonian that is formally integrable,
since it is expressed only in terms of the actions $R_j$, $j=1,\ldots, n$.
For more theoretical details we refer the reader to~\cite{arnold1} and \cite{simo}, and for practical
implementations, the works of~\cite{machuy}  and~\cite{stuchi}.
In this paragraph we have tried to stress a point that is usually not made clearly when calculating the
Birkhoff normal forms for the photogravitational problem: the difference between the natural resonances of the $2n$ complex eigenvalues,
and the resonances of the real eigenfrequencies $\omega_1$ and $\omega_2$.

Perhaps due to Hamilton-Jacobi tradition, some engineers and physicists, for example  Kumar and Choudhry~\cite{ku},
prefer to work with real variables and mixed generating functions which depend on both old and new variables.
In this work  we use the Lie derivative approach, which allows us to use a generating function dependent only on new the variables,
with the advantage of turning the cumbersome passage to the normal formal more transparent.
Moreover, with this approach, when the coefficients are not algebraic, but numerical, one can write fast codes to construct very high order normal forms
(see for example \cite{simo}, \cite{stuchi}).

Before describing the algorithm and the form of the generating function, a further step is necessary to prepare the Hamiltonian for the process of
normalization. It is easier to perform the manipulations in the complex field and, to this end, we shall make use of
the change of variables, ${\bf R_C}$, and its inverse, ${\bf C_R}$:
\begin{eqnarray*}\label{trans-co}
 {\bf R_C:} & \quad x_j =\frac{1}{\sqrt{2}}(\bar{\xi}_j-{\bf i}\bar{\eta}_j), & \quad y_j= \tfrac{1}{\sqrt{2}}(-{\bf i}
 \bar{\xi}_j+\bar{\eta}_j), \\
{\bf C_R:} & \quad \bar{\xi}_j =\frac{1}{\sqrt{2}}(x_j+{\bf i}y_j),   & \quad \bar{\eta}_j= \tfrac{1}{\sqrt{2}}({\bf i} x_j+y_j),
\end{eqnarray*}
where ${\bf i}=\sqrt{-1}$, to transform  $\bar H(\bar \xi_1,\bar \xi_2,\bar \eta _1,\bar \eta_2)$ into the complex Hamiltonian
\begin{equation}\label{hamco}
  H(x_1,x_2,y_1,y_2)={\bf i}\: \omega_1 x_1 y _1 +{\bf i}\: \omega_2 x_2y_2 +\!\!\!\!\!\!\!\! \sum_{k_1+k_2+l_1+l_2=3}\!\!\!\!\!\!\!\! h_{k_1 k_2 l_1 l_2}\;x_1^{k_1}x_2^{k_2}y_1^{l_1}y_2^{l_2}.
\end{equation}
In fact, the complexification can be done directly from Hamiltonian~(\ref{ham2}) to~(\ref{hamco}), as is the case in this work (Appendix~B).

Now, we introduce the generating function $G(x_1,x_2,y_1,y_2)$ as a series expansion in homogeneous polynomials of degree $m\ge 3$
\begin{equation}\label{GF}
G_m(x_1,x_2,y_1,y_2)=\sum_{j=3}^m G_j, \quad \mbox{where} \quad G_j=\!\!\!\!\!\!\!\!\sum_{k_1+k_2+l_1+l_2=j} \!\!\!\!\!\!\!\!g_{k_1k_2l_1l_2}\;x_1^{k_1}x_2^{k_2}y_1^{l_1}y_2^{l_2},
\end{equation}
and the associated canonical transformation $T_G$,  such that
$$T_G\;H(x_1,x_2,y_1,y_2)=Z(X_1,X_2,Y_1,Y_2)=H(T_G(x_1,x_2,y_1,y_2)),$$
with
\begin{equation}\label{lieseries}
Z(X_1,X_2,Y_1,Y_2)=H+\{H,G\}+\frac{1}{2!}\{\{H,G\},G\}+\frac{1}{3!}\{\{\{H,G\},G\},G\}+\cdots,
\end{equation}
where $\{H,G\}=L_HG$ is the usual Poisson bracket (or Lie derivative) of the functions $H$ and $G$. This means that $Z$ is the time one flow of the flux
generated by the canonical system of equations associated with the generating function $G$. To find the Birkhoff
normal form and the corresponding change of variables (and its inverse) one has to determine the generating function that gives
the prescribed form discussed above. Collecting powers of~(\ref{lieseries}) up to order four gives
\begin{align}
Z_2=& H_2, \nonumber \\
Z_3=& H_3+\{H_2,G_3\},  \label{zo4} \\
Z_4=& H_4+\{H_3,G_3\}+\frac{1}{2!}\{\{H_2,G_3\},G_3\}+\{H_2,G_4\}.\nonumber
\end{align}

Since we are supposing that there are no resonances besides the natural ones (as explained above),  all terms of order 3 can be removed from~(\ref{lieseries}) by setting
$$\{H_2,G_3\}=Z_3-H_3.$$
Solving for $Z_3=0$ gives
\begin{equation}\label{g3}
G_3=\sum_{k_1+k_2+l_1+l_2=3}\displaystyle \frac{h_{k_1 k_2 l_1 l_2}}{({\bf k}-{\bf l},\Lambda)}\;x_1^{k_1}x_2^{k_2}y_1^{l_1}y_2^{l_2},
\end{equation}
where
$$({\bf k}-{\bf l},\Lambda)={\bf i}(k_1-l_1)\omega_1+{\bf i}(k_2-l_2)\omega_2 $$
are the coefficients of the monomials of the Poisson bracket $\{H_2,G_3\}$, which are nonzero outside the resonance regions.

For terms of order four, the solution is not so easy since $Z_4 \ne 0$, and we have to solve simultaneously for $G_4$ and
$Z_4$, and so on, up to the required order, $m$, of the normal form. In this case
one can use the general result~\cite{simo} for $k \ge 3$:
\begin{equation}\label{homo}
\{H_2,G_k\}+Z_k=F_k,   \qquad F_3=H_3,\; Z_2=H_2
\end{equation}
and
\begin{equation}\label{fk}
F_k= \sum_{m=1,\ldots,k-3}\frac{m}{k-2}\{G_{2+m},Z_{k-m}\}+\sum_{m=1,\ldots,k-2}\frac{m}{k-2}H_{2+m,k-m-2}
\end{equation}
where
\begin{equation}\label{hinterm}
H_{2+m,k-m-2}=\sum_{j=1}^{k-m-2}\frac{j}{k-m-2}\{G_{2+j},H_{2+m,k-j-m-2}\}+H_k.
\end{equation}

Applying these formulae we can find $Z_4$ and $G_4$, which give us the fourth order terms that the Arnold-Moser Theorem requires.
The algebraic manipulations described so far were performed using
the software Maple, but any algebraic software could have been used. Note that $Z_4$ is a complex Birkhoff normal form but, using the transformation
$C_R$ given in~(\ref{trans-co}), we obtain finally
\begin{align}
& Hr_2 = \frac{\omega_1}{2}(\xi_1^2 + \eta _1^2) -\frac{\omega_2}{2}(\xi_2^2 +\eta _2^2), \nonumber\\
& Hr_3= 0, \label{rnoform} \\
& Hr_4= \delta_{11}(\xi_1^2+\eta_1^2)^2+\delta_{12}(\xi_1^2+\eta_1^2)(\xi_2^2+\eta_2^2)+\delta_{22}(\xi_2^2+\eta_2^2)^2, \nonumber
\end{align}
which is the form required by the Arnold-Moser Theorem.

According to this theorem we have to check whether the determinant
\begin{equation}\label{det}
D(\kappa_1,\kappa_2) = -(\delta_{11}\,\omega_2^2-2\delta_{12}\,\omega_1\omega_2+\delta_{22}\,\omega_1^2)
\end{equation}
is nonzero.
If $D(\kappa_1,\kappa_2) \neq 0$ for some pair $(\kappa_1, \kappa_2)$ then, for this pair, the motion is stable in the Lyapunov sense
provided $\omega_1\neq \omega_2$, $\omega_1\neq 2\omega_2$ or $\omega_1\neq 3\omega_2$.

The complex canonical transformation that takes the Taylor series~(\ref{ham2}) to~(\ref{hamco}) is shown in Appendix~B while the coefficients
of the fourth order normal form are given in Appendix~C. After a sequence of algebraic manipulations we finally obtain
the determinant as a rational function, whose numerator is a quintic in $\mu$.  Explicitly,
defining the product of factors
\[
\Pi\equiv \left( \kappa_1+1+\kappa_2 \right)  \left( \kappa_1+1-
\kappa_2 \right)  \left( \kappa_1-1+\kappa_2 \right)
 \left( \kappa_1-1-\kappa_2 \right),
\]
we can write $D$ as
\[
D(\kappa_1,\kappa_2) = \frac{1}{X} \sum_{i=0}^5 D_i\,\mu^{i}
\]
where the denominator, $X$, is
\begin{eqnarray*}
X = 2048\,\Pi {\kappa_1}^{4}{\kappa_
{{2}}}^{4} (225\, \Pi {\mu}(\mu-1)-16\,{\kappa_1}^{2}{\kappa_2}^{
2})
\left(
9\,\Pi \mu(\mu-1)-{\kappa_1}^{2}{\kappa_2}^{2}
\right)
\end{eqnarray*}
and the coefficients, $D_i$, of the quintic in the numerator are given by
\begin{eqnarray*}
\frac{D_5}{12960 \Pi^3} & = &
 (\kappa_1^2-\kappa_2^2)
( 5\,{\kappa_1}^{2}-5-8\,\kappa_1\kappa_2+5\,{\kappa_2}^{2} )
( 5\,{\kappa_1}^{2}-5+8\,\kappa_1\kappa_2+5\,{\kappa_2}^{2} ),
\end{eqnarray*}
\begin{eqnarray*}
-\frac{D_4}{2592\Pi^2} & = &
 375\,{\kappa_1}^{2}
-1500\,{\kappa_1}^{4}
+2250\,{\kappa_1}^{6}
-1500\,{\kappa_1}^{8}
+375\,{\kappa_1}^{10}
\\
& &
-250\,{\kappa_2}^{2}
+1000\,{\kappa_2}^{4}
-1500\,{\kappa_2}^{6}
+1000\,{\kappa_2}^{8}
-250\,{\kappa_2}^{10}\\
& &
-350\,{\kappa_1}^{2}{\kappa_2}^{2}
+500\,{\kappa_1}^{2}{\kappa_2}^{4}
-1450\,{\kappa_1}^{2}{\kappa_2}^{6}
+925\,{\kappa_1}^{2}{\kappa_2}^{8}\\
& &
+{\kappa_1}^{4}(150\,{\kappa_2}^{2}
+8\,{\kappa_2}^{4}
-1650\,{\kappa_2}^{6})
+1750\,{\kappa_1}^{6}{\kappa_2}^{2}
+1900\,{\kappa_1}^{6}{\kappa_2}^{4}\\
& &
-1300\,{\kappa_1}^{8}{\kappa_2}^{2},
\end{eqnarray*}
\begin{eqnarray*}
\frac{D_3}{288 \Pi^2} &=&
3375\,{\kappa_1}^{2}
-13500\,{\kappa_1}^{4}
+20250\,{\kappa_1}^{6}
-13500\,{\kappa_1}^{8}
+3375\,{\kappa_1}^{10}\\
& &
-1125\,{\kappa_2}^{2}
+4500\,{\kappa_2}^{4}
-6750\,{\kappa_2}^{6}
+4500\,{\kappa_2}^{8}
-1125\,{\kappa_2}^{10}\\
& &
-6300\,{\kappa_1}^{2}{\kappa_2}^{2}
+7495\,{\kappa_1}^{2}{\kappa_2}^{4}
-9590\,{\kappa_1}^{2}{\kappa_2}^{6}
+5020\,{\kappa_1}^{2}{\kappa_2}^{8}
\\
& &
+4205\,{\kappa_1}^{4}{\kappa_2}^{2}
+144\,{\kappa_1}^{4}{\kappa_2}^{4}
-10065\,{\kappa_1}^{4}{\kappa_2}^{6}
\\
& &
+14990\,{\kappa_1}^{6}{\kappa_2}^{2}
+14565\,{\kappa_1}^{6}{\kappa_2}^{4}
-11770\,{\kappa_1}^{8}{\kappa_2}^{2}
\end{eqnarray*}
\begin{eqnarray*}
\frac{-D_2}{288\,{\kappa_1}^{2} \Pi} &=&
1125
-6750{\kappa_1}^{2}
+16875{\kappa_1}^{4}
-22500{\kappa_1}^{6}
+16875{\kappa_1}^{8}
-6750{\kappa_1}^{10}\\
&&+1125\,{\kappa_1}^{12}
-5400\,{\kappa_2}^{2}
+11050\,{\kappa_2}^{4}
-12700\,{\kappa_2}^{6}
+8925\,{\kappa_2}^{8}\\
&&
-3700{\kappa_2}^{10}
+700{\kappa_2}^{12}
+14890{\kappa_1}^{2}{\kappa_2}^{2}
-9518{\kappa_1}^{2}{\kappa_2}^{4}
-1634{\kappa_1}^{2}{\kappa_2}^{6}\\
& &
+6012\,{\kappa_1}^{2}{\kappa_2}^{8}
-3000\,{\kappa_1}^{2}{\kappa_2}^{10}
-5560\,{\kappa_1}^{4}{\kappa_2}^{2}
-674\,{\kappa_1}^{4}{\kappa_2}^{4}\\
&&
-3712\,{\kappa_1}^{4}{\kappa_2}^{6}
+7295\,{\kappa_1}^{4}{\kappa_2}^{8}
-18660\,{\kappa_1}^{6}{\kappa_2}^{2}
-14298\,{\kappa_1}^{6}{\kappa_2}^{4}\\
&&-12850\,{\kappa_1}^{6}{\kappa_2}^{6}
+13440\,{\kappa_1}^{8}{\kappa_2}^{4}
+21440\,{\kappa_1}^{8}{\kappa_2}^{2}
-6710\,{\kappa_1}^{10}{\kappa_2}^{2}
\end{eqnarray*}
\begin{eqnarray*}
\frac{-D_1}{32\,{\kappa_1}^{4}{\kappa_2}^{2} \Pi}&=&
3465
-13860\,{\kappa_1}^{2}
+20790\,{\kappa_1}^{4}
-13860\,{\kappa_1}^{6}
+3465\,{\kappa_1}^{8}\\
& &
-2790\,{\kappa_2}^{2}
-4780\,{\kappa_2}^{4}
+4070\,{\kappa_2}^{6}
+35\,{\kappa_2}^{8}\\
& &
-2870\,{\kappa_1}^{6}{\kappa_2}^{2}
-5196\,{\kappa_1}^{4}{\kappa_2}^{4}
+2950\,{\kappa_1}^{4}{\kappa_2}^{2}\\
& &
+2710\,{\kappa_1}^{2}{\kappa_2}^{2}
+4566\,{\kappa_1}^{2}{\kappa_2}^{6}
+15912\,{\kappa_1}^{2}{\kappa_2}^{4}
\end{eqnarray*}
\begin{eqnarray*}
\frac{-D_0}{512\,{\kappa_1}^{6}{\kappa_2}^{4}}&=&
5
-20\,{\kappa_1}^{2}
+30\,{\kappa_1}^{4}
-20\,{\kappa_1}^{6}
+5\,{\kappa_1}^{8}
+10\,{\kappa_2}^{2}
-36\,{\kappa_2}^{4}
+22\,{\kappa_2}^{6}\\
&&-{\kappa_2}^{8}
-10\,{\kappa_1}^{2}{\kappa_2}^{2}
+56\,{\kappa_1}^{2}{\kappa_2}^{4}
+22\,{\kappa_1}^{2}{\kappa_2}^{6}
-10\,{\kappa_1}^{4}{\kappa_2}^{2}\\
&&
-36\,{\kappa_1}^{4}{\kappa_2}^{4}
+10\,{\kappa_1}^{6}{\kappa_2}^{2}
\end{eqnarray*}

\begin{figure}[ht]
\label{Dzero}
\begin{center}
\includegraphics[height=80mm]{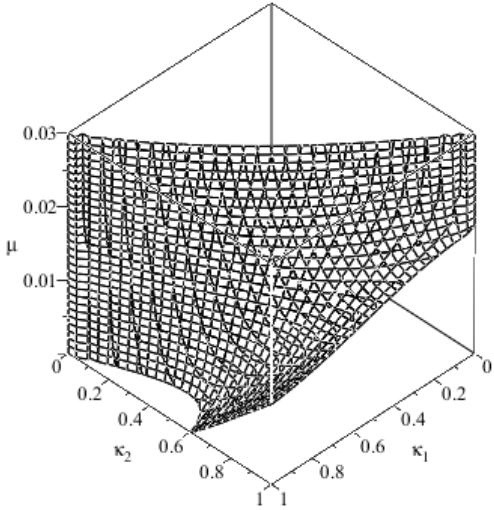}
\end{center}
\caption{The surface $D(\mu,\kappa_1,\kappa_2)=0$.}
\end{figure}

Since the numerator of $D$ is a polynomial of order 5, it is, in general, impossible to write an
algebraic expression for the values for which $\mu(\kappa_1,\kappa_2)$ produces $D=0$.
However it is easy to plot the surface $D=0$, which is shown in figure~\ref{Dzero} for the
parameter ranges $\kappa_1\in[0,1]$, $\kappa_2\in[0,1]$, $\mu\in[0,0.3]$.

\begin{figure}[ht]
\label{figDeprit}
\begin{center}
\includegraphics[height=50mm]{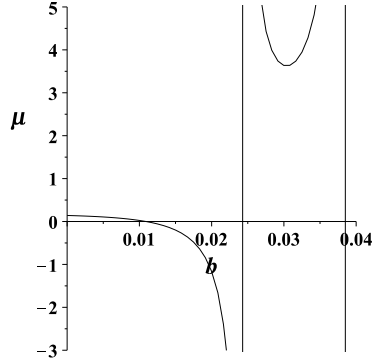}
\end{center}
\caption{The classical determinant, $D_c$, for $0 \leq \mu\leq 0.04$}
\end{figure}

This surface seems to be equivalent to that shown by Go\'zdziewski et. al.~\cite{gmn}, who use slightly different variables. They, however, do not
give an explicit expression for the determinant (their $D_4$), while our expression allows a better understanding of the structure of this function.
For example it is possible to see that, for certain combinations of $\kappa_1$ and $\kappa_2$,
the numerator of $D$ reduces to a quartic, and an explicit algebraic expression for
$\mu(\kappa_1,\kappa_2)$
on the surface $D=0$ can be obtained.
In the physical region of the parameter space this happens when
\[
\kappa_1+\kappa_2 =1, \qquad \kappa_1=\kappa_2,~~~{\rm or}~~~~5(\kappa_1^2+\kappa_2^2-1)+8\kappa_1\kappa_2=0.
\]
The second of these is particularly interesting, since it
includes the purely gravitational case $\kappa_1=\kappa_2=1$. For these values we have
\[
D=D_c = {\frac {9}{64}}\,{\frac {13041\,{\mu}^{4}-26082\,{\mu}^{3}+14664\,{
\mu}^{2}-1623\,\mu+16}{ \left( 675\,{\mu}^{2}-675\,\mu+16 \right)
 \left( 27\,{\mu}^{2}-27\,\mu+1 \right) }},
\]
with the first positive root at
\[
\mu=\frac{1}{2}-{\frac {1}{2898}}\,\sqrt {1576995+966\,\sqrt {199945}} \approx 0.0109136676
\]
in agreement with the classical result of Deprit and Deprit-Bartholom\'{e}. For illustration
we show the behavior of $D_c(\mu)$ in Figure~\ref{figDeprit}.

\section{Conclusion}

One of the motivations of this paper was to bring together various results relating to the restricted photogravitational three-body problem scattered about the literature,
with special emphasis on the stability of the $L_4$ and $L_5$ libration points in the absence of first and second-order resonances.
As part of this work, we noticed that, though the fourth-order Taylor series expansion given by Kumar and Choudhry is correct,
one of their results apparently does not reproduce the classical case in the appropriate limits.
This motivated us to use a different approach, namely the Lie triangle method,  to calculate the Birkhoff normal form for the Hamiltonian.
Along the way,  it was found that the conditions for the existence of resonances given by Kumar and Choudhry had to be slightly modified.

In this paper we give explicitly the fourth-order normal form, as well as the complex canonical transformation used to prepare the Hamiltonian
for the Birkhoff normalization. We also provide an explicit expression in terms of a rational function, for the fourth-order determinant, $D(\mu,\kappa_1,\kappa_2)$,
and show that it reduces to the classical, purely gravitational case in the limits $\kappa_1 = \kappa_2\rightarrow 1$.

Our algebraic analysis seems to corroborate the numerical fourth-order treatment by Go\'zdziewski et. al.~\cite{gmn}.
The algebraic treatment shows that the surface $D=0$ is, in fact, generated by a polynomial that is fifth-order in $\mu$. Knowing the explicit behavior of $D$ means that special cases can be
studied, and we find three relations in the physical parameter space of
$\kappa_1$ and $\kappa_2$ for which the numerator reduces to a quartic.
In these cases explicit expressions for the surface $D=0$ can therefore be obtained. One
of the cases, $\kappa_1=\kappa_2$,  includes the classical non-radiational case and we show
how the classical result of Deprit and Deprit-Bartholom\'{e} follows as a special case.

\newpage

\section*{Appendix A}
\noindent The coefficients of the Taylor series  expansion for $H_3$ and $H_4$.
$h_{ijkl}$ is the coefficient of $q_1^i\,q_2^j \, p_1^k \,p_2^l$ in the Hamiltonian.

\begin{align*}
h_{3000} = &\frac{\mu}{16 \kappa_2^4}\,(\kappa_1^2-1-\kappa_2^2)\left[5 (\kappa_1^2-1-\kappa_2^2)^2-12\kappa_2^2\right]\\
& \qquad + \frac{1-\mu}{16\kappa_1^4}\, (\kappa_1^2+1-\kappa_2^2)\left[5 (\kappa_1^2+1-\kappa_2^2)^2-12\kappa_1^2\right],\\
h_{2100} = & \frac{3}{8}\:  \frac{\kappa_1}{\kappa^3_2}\: \sqrt{b}\:  \mu\:\left[ 5 (\kappa_1^2-1-\kappa_2^2)^2-4\kappa_2^2\right]\\
& \qquad + \frac{3}{8}\: \frac{\kappa_2}{\kappa^3_1}\:\sqrt{b}\:(1-\mu)\left[5 (\kappa_1^2+1-\kappa_2^2)^2-4\kappa_1^2\right],\\
h_{1200} = & \frac{3\mu}{4\kappa_2^2}\,(\kappa_1^2-1-\kappa_2^2)(5b\kappa_1^2-1) +  \frac{3(1-\mu)}{4\kappa_1^2}\, (\kappa_1^2+1-\kappa_2^2)(5b\kappa_2^2-1),\\
h_{0300} = & \frac{\kappa_1}{2\kappa_2} \sqrt{b}  \:\mu(5b\kappa_1^2-3) +  \frac{\kappa_2}{2\kappa_1}\sqrt{b}\:
 \:(1-\mu)(5b\kappa_2^2-3),
\end{align*}
%
%
\begin{align*}
h_{4000} = & -\frac{1-\mu}{8\kappa_1^6} \left[3\kappa_1^4-\frac{15}{2}(\kappa_1^2+1-\kappa_2^2)^ 2\kappa_1^2+\frac{35}{16}(\kappa_1^2+1-\kappa_2^2)^4\right]
\\
& \qquad - \frac{\mu}{8 \kappa_2^6} \left[ 3\kappa_2^4-\frac{15}{2}(\kappa_1^2-1-\kappa_2^2)^2\kappa_2^2+\frac{35}{16}(\kappa_1^2-1-\kappa_2^2)^4\right], \\
h_{3100} = & \frac{5 \sqrt{b}}{4} \: \frac{\kappa_2}{\kappa_1^3} (1-\mu)(\kappa_1^2+1-\kappa_2^2)\left[3- \frac{7(\kappa_1^2+1-\kappa_2^2)^2}{4\kappa_1^2} \right]\\
 & \qquad + \frac{5 \sqrt{b}}{4} \frac{\kappa_1}{\kappa_2^3} \mu (\kappa_1^2-1-\kappa_2^2)\left[3- \frac{7(\kappa_1^2-1-\kappa_2^2)^2}{4\kappa_2^2}\right],\\
h_{2200} = &  \frac{3(1-\mu)}{4 \kappa_1^2} \, \left[5\kappa_2^2b + \frac{5}{4\kappa_1^2}(\kappa_1^2+1-\kappa_2^2)^2
-\frac{35}{4\kappa_1^2}\: b\kappa_2^2(\kappa_1^2+1-\kappa_2^2)^2-1\right] \\
& \quad + \frac{3\mu}{4 \kappa_2^2} \,\left[5\kappa_1^2b + \frac{5}{4\kappa_2^2}(\kappa_1^2-1-\kappa_2^2)^2
-\frac{35}{4\kappa_2^2}\: b\kappa_1^2(\kappa_1^2-1-\kappa_2^2)^2-1\right], \\
h_{1300} = & \frac{5}{4\kappa_1^3}\: \kappa_2\sqrt{b} \: (1-\mu)\:(\kappa_1^2+1-\kappa_2^2)(3-7\kappa_2^2b)\\
& \qquad + \frac{5}{4\kappa_2^3}\: \kappa_1\sqrt{b}\: \mu\: (\kappa_1^2-1-\kappa_2^2)(3-7\kappa_1^2b) . \\
h_{0400} = & -\frac{1}{8 \kappa_1^2} (1-\mu) \left(3-30\kappa_2^2b+35\kappa_2^4b^2\right) -\frac{1}{8 \kappa_2^2} \:\mu \left(3-30\kappa_1^2b+35\kappa_1^4b^2\right).
\end{align*}

\section*{Appendix B}
\noindent The canonical transformation from the variables $(q_i,p_i)$ to $(Q_i,P_i)$.
\[
q_1 = \frac{-1+i}{\sqrt{2}}\left\{\frac{(B-2i\omega_1)Q_1+(B+2i\omega_1)P_1
}{\sqrt{\omega_1(2\omega_1^2-1)(-3+4A-2\omega_1^2)}}
 +\frac{(B-2i\omega_2)Q_2+(B+2i\omega_2)P_2
}{\sqrt{\omega_2(2\omega_2^2-1)(-3+4A-2\omega_2^2)}}
\right\}
\]
\[
q_2 = \frac{-1+i}{2\sqrt{2}}
\left\{
\left[\frac{-3+4A-2\omega_1^2}{\omega_1(2\omega_1^2-1)}\right]^{1/2}(Q_1+P_1)
+\left[\frac{-3+4A-2\omega_2^2}{\omega_2(2\omega_2^2-1)}\right]^{1/2}(Q_2+P_2)
\right\}
\]
\begin{eqnarray*}
p_1 & = & \frac{-1+i}{2\sqrt{2}}
\left\{
\frac{(2\omega_1^2+2iB\omega_1+4A-3)Q_1+(2\omega_1^2-2iB\omega_1+4A-3)P_1}
{\sqrt{\omega_1(2\omega_1^2-1)(-3+4A-2\omega_1^2)}}\right.\\
&& \left. +
\frac{(2\omega_2^2+2iB\omega_2+4A-3)Q_2+(2\omega_2^2-2iB\omega_2+4A-3)P_2}
{\sqrt{\omega_2(2\omega_2^2-1)(-3+4A-2\omega_2^2)}}
\right\}
\end{eqnarray*}
\begin{eqnarray*}
p_2  & = &
 \frac{-1+i}{2\sqrt{2}} \left\{
\frac{[-2i\omega_1^3+(4A+1)\omega_1-2B] Q_1+[-2i\omega_1^3+(4A+1)\omega_1+2B] P_1}{\sqrt{\omega_1(2\omega_1^2-1)(-3+4A-2\omega_1^2)}}\right.\\
&&\left.
+
\frac{[-2i\omega_2^3+(4A+1)\omega_2-2B] Q_2+[-2i\omega_2^3+(4A+1)\omega_2+2B] P_2}{\sqrt{\omega_2(2\omega_2^2-1)(-3+4A-2\omega_2^2)}}
\right\},
\end{eqnarray*}
\noindent
where $A$ and $B$ are as defined in~(\ref{ABdef}), with $\omega_1$ and $\omega_2$ as in~(\ref{frec}).

\section*{Appendix C}\noindent Applying  the transformation
of Appendix~B to the Taylor series given in Appendix~A and applying the normal form procedure described in section 4.2, we obtain the
following expressions for the coefficients 
of the fourth-order Birkhoff normal after converting to real variables variables  $(x_1,x_2,y_1,y_2)$:
\begin{eqnarray*}
\delta_{11}&=&
\frac{1}{4}\left(-{\frac {\,iHQP_{1011}HQP_{1110}}{\omega_2}}
-{\frac {\,iHQP_{2001}HQP_{0120}}{2(-2\,\omega_1+\omega_2)}}\right.\\
&&
 +{\frac {\,iHQP_{2001}{HQP}_{0120}}{2(2\,\omega_1-\omega_2)}}-
\,HQP_{2020}-{\frac {3\,iHQP_{2010}HQP_{1020}}{\omega_1}}\\
&&-{\frac {\,iHQP_{0021}HQP_{2100}}{(
\,\omega_1+\omega_2)}}-{\frac {3\,iHQP_{0030}{HQP}_{3000}}{\omega_1}}\\
&&\left.+{\frac {\,iHQP_{0021}{HQP}_{2100}}{2(-2\,\omega_1-\omega_2)}}\right)
\end{eqnarray*}
\begin{eqnarray*}
\delta_{22}&=&
  -\frac{1}{4}\,HQP_{1111}+
\frac{1}{2}\left(-{\frac {\,iHQP_{0012}{HQP}_{1200}}{\omega_1+2\,\omega_2}}
-{\frac {\,i{HQP}_{{0102}}HQP_{1110}}{\omega_2}}\right.\\
&&-{\frac {\,i{HQP}_{{0201}}HQP_{1011}}{\omega_2}}
+{\frac {\,iHQP_{0210}HQP_{1002}}{\omega_1
-2\,\omega_2}}\\
&&+{\frac {\,iHQP_{0012}HQP_{1200}}{-\omega_1-2\,\omega_2}}-{\frac {\,iHQP_{0210}{HQP}_{1002}}{-\omega_1+2\,\omega_2}}\\
&&-{\frac {\,i{HQP}_{1020}HQP_{{1101}}}{\omega_1}}-{\frac {\,i{HQP}_{0021}HQP_{2100}}{2\,\omega_1+{\omega_2}}}\\
&&-{\frac {\,iHQP_{{0111}}HQP_{2010}}{\omega_1}}-{\frac {\,iHQP_{0120}HQP_{2001}}{2\,\omega_1-\omega_2}}\\
&&\left.+{\frac {\,iHQP_{0120}{HQP}_{2001}}{-2\,\omega_1+\omega_2}}+{\frac {\,i{HQP}_{0021}HQP_{2100}}{-2\,\omega_1-\omega_2}}\right)
\end{eqnarray*}
\begin{eqnarray*}
\delta_{12}&=&
-\frac{1}{4}\,HQP_{1111}+\frac{1}{2}\left({\frac {\,iHQP_{0012}{HQP}_{1200}}{\omega_1+2\,\omega_2}}-{\frac {\,i{HQP}_{{0102}}HQP_{1110}}{\omega_2}}\right.\\
&&-{\frac {\,i{HQP}_{{0201}}HQP_{1011}}{\omega_2}}+{\frac {
\,iHQP_{0210}HQP_{1002}}{\omega_1-2\,\omega_2}}\\
&&+{\frac {\,iHQP_{0012}HQP_{1200}}{-{
\omega_1}-2\,\omega_2}}
-{\frac {\,iHQP_{0210}{HQP}_{1002}}{-\omega_1+2\,\omega_2}}\\
&&-{\frac {\,i{HQP}_{1020}HQP_{{1101}}}{\omega_1}}
-{\frac {\,i{HQP}_{0021}HQP_{2100}}{2\,\omega_1+\omega_2}}\\
&&-{\frac {\,iHQP_{{0111}}HQP_{2010}}{\omega_1}}-{\frac {\,iHQP_{0120}HQP_{2001}}{2
\,\omega_1-\omega_2}}\\
&&\left.+{\frac {\,iHQP_{0120}{HQP}_{2001}}{-2\,\omega_1+\omega_2}}+{\frac {\,i{HQP}_{0021}HQP_{2100}}{-2\,\omega_1-\omega_2}} \right)
\end{eqnarray*}

In the complex form we have
\[
H(Q_i,P_i) = \Delta_{11} \,P_1^2 Q_1^2 + \Delta_{22}\, P_2^2 Q_2^2 + \Delta_{12}\, P_1 P_2 Q_1 Q_2
\]
where
\begin{eqnarray*}
\Delta_{11}  & = &
HQP_{2020}
+\frac{3i}{\omega_1}(HQP_{1020}HQP_{2010}+HQP_{0030}HQP_{3000})\\
&&+{\frac {iHQP_{1110}HQP_{1011}}{\omega_2}}
-{\frac {iHQP_{2001}HQP_{0120}}{2\,\omega_1-\omega_2}}
+{\frac {iHQP_{0021}HQP_{2100}}{2\,\omega_1+\omega_2}},
\end{eqnarray*}
\begin{eqnarray*}
\Delta_{12} & = &
HQP_{1111}
+{\frac {4\,iHQP_{0012}HQP_{1200}}{\omega_1+2\,\omega_2}}
-{\frac {4\,iHQP_{0210}HQP_{1002}}{\omega_1-2\,\omega_2}}\\
& &
+{\frac {4\,iHQP_{0021}HQP_{2100}}{2\,\omega_1+\omega_2}}
+{\frac {4\,iHQP_{0120}HQP_{2001}}{2\,\omega_1-\omega_2}}\\
&&+\frac{2\,i}{\omega_1}\,(HQP_{0111}\,HQP_{2010}+HQP_{1101}\,HQP_{1020})\\
&&+\frac{2\,i}{\omega_2}\,(HQP_{0102}\,HQP_{1110}+HQP_{0201}\,HQP_{1011})
\end{eqnarray*}
and $\Delta_{22}$ can be obtained from $\Delta_{11}$ by making the changes $\omega_1\leftrightarrow\omega_2$
and $HQP_{ijkl}\rightarrow HQP_{lkji}$.

\section*{Acknowledgments}

M. Alvarez-Ram\'{\i}rez has been partially supported by Red de cuerpos acad\'emicos Ecuaciones Diferenciales, Proyecto Sistemas din\'amicos y estabilizaci\'on. PROMEP 2011-SEP M\'exico. R. V. de Moraes and T. J. Stuchi have been supported by a grant from CNPq, Brazil. R.V. de Moraes  also acknowledges partial financial support from CAPES.

\end{document}